\documentclass[twocolumn,amsmath,amssymb,prd,reprint,longbibliography,floatfix,8pt]{revtex4-1}

\usepackage{cancel}
\usepackage{enumitem}
\usepackage[utf8x]{inputenc}
\usepackage{graphicx}
\usepackage{dcolumn}
\usepackage{bm}
\usepackage{hyperref}
\usepackage[switch]{lineno}
\usepackage{float}             
\usepackage{subfigure}
\usepackage{tikz}
\usetikzlibrary{arrows.meta}

\begin{document}

\title{Towards exact predictions of spin-phonon relaxation times: an \textit{ab initio} implementation of open quantum systems theory}

\author{Alessandro Lunghi}
\email{lunghia@tcd.ie}
\affiliation{School of Physics, AMBER and CRANN Institute, Trinity College, Dublin 2, Ireland}

\begin{abstract}
{\bf Spin-phonon coupling is the main drive of spin relaxation and decoherence in solid-state semiconductors at finite temperature. Controlling this interaction is a central problem for many disciplines, ranging from magnetic resonance to quantum technologies. Spin relaxation theories have been developed for almost a century but often employ a phenomenological description of phonons and their coupling to spin, resulting in a non-predictive tool and hindering our detailed understanding of spin dynamics. Here we combine fourth-order time-local quantum master equations with advanced electronic structure methods and perform predictions of spin-phonon relaxation time for a series of solid-state coordination compounds based on both transition metals and lanthanide Kramers ions. The agreement between experiments and simulations demonstrates that an accurate, universal and fully \textit{ab initio} implementation of spin relaxation theory is possible, thus paving the way to a systematic study of spin-phonon relaxation in solid-state materials.}
\end{abstract}

\maketitle

Relaxation of localized magnetic moments in condensed-phase is mainly driven by three interactions: spin-spin, spin-conduction electron and spin-phonon coupling\cite{wu2010spin}. In the case of semiconductors at temperatures above a few K, the latter is the most important one. In these circumstances, the oscillatory motion of atoms acts as a time-dependent perturbation on the spin degrees of freedom, due to the presence of spin-orbit and spin-spin interactions, and leads to spin thermalization and loss of spin coherence. On the one hand a deep understanding of spin-phonon relaxation and spin-phonon coupling has the potential to open up new ways to address central problems for modern technologies such as magnetic information storage\cite{maehrlein2018dissecting}, spintronics\cite{vzutic2004spintronics}, quantum information science\cite{wolfowicz2021quantum,atzori2019second}, magnetic resonance\cite{eaton2002relaxation}, and more. On the other hand, the study of spin relaxation in condensed-phase represents a fundamental benchmark for our understanding of the theory of open quantum systems. Although a formally exact description of the quantum dynamics of open systems is achievable through time-convolutionless\cite{breuer2001time,timm2011time} or Nakajima-Zwanzig\cite{koller2010density} equations, their application is often done either in a parametric fashion or for simplified models, therefore escaping the ultimate benchmark of a full \textit{ab initio} implementation for realistic systems. Spin-phonon coupling represents a paradigmatic example of open-quantum-system dynamics and offers a unique test-bed for open quantum system theory and \textit{ab initio} simulations at the same time.

Spin-phonon relaxation has now been debated for almost a century starting with the seminal works of Waller\cite{waller1932magnetisierung}, Van Vleck\cite{van1940paramagnetic}, Redfield\cite{redfield1957theory} and Orbach\cite{orbach1961spin}. Redfield relaxation theory and its adaptations form the fundamental building block of our understanding of spin relaxation and have served as the basis to model both nuclear and electronic spin relaxation experiments. Redfield theory is generally adapted to the case of solid-state systems by accounting for a Debye-like phonon density of state and by including all the details about spin-phonon coupling under a single average phenomenological coupling constant\cite{orbach1961spin}. Despite the incredible usefulness of such a formalism for interpreting experiments, these approximations have the major drawback of making it impossible to use the theory as a real predictive tool. 
\begin{figure*}[t]
 \begin{center}
    \includegraphics[scale=1]{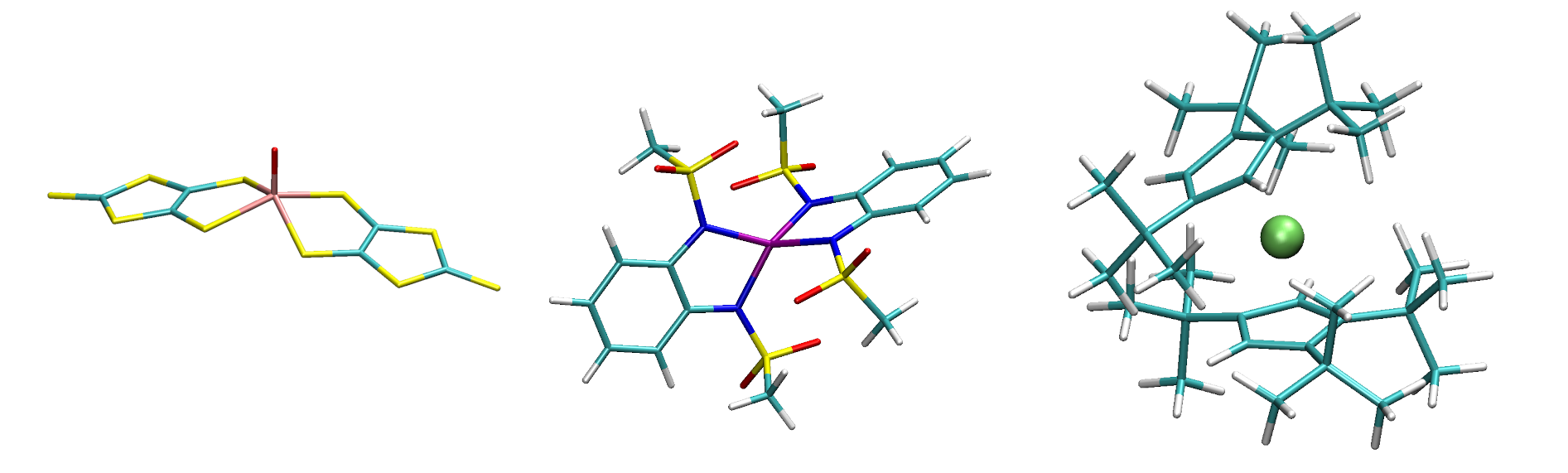}
\end{center}
 \caption{\textbf{Molecular structures}. Left, central and right panels report the structures of \textbf{1},  \textbf{2} and \textbf{3}, respectively. Color code: pink for Vanadium, Purple for Cobalt, light green for Dysprosium, green for Carbon, blue for Nitrogen, red for Oxygen, yellow for Sulphur and white for Hydrogen.}
 \label{geos}
\end{figure*}
Only recently, fully-\textit{ab initio} methods to predict spin-phonon relaxation in solid-state materials has been proposed\cite{escalera2017determining,lunghi2017role,goodwin2017molecular,lunghi2019phonons,park2020spin,xu2020spin}. At the heart of these methods there is a combination of Redfield theory of relaxation and electronic structure methods. The latter are used to define all the coefficients needed to set up the former on a material-specific case and thus enable a non-parametric treatment of all the relevant interactions, such as spin and phonon spectra, and all the spin-phonon coupling coefficients. Applications of this method to magnetic molecules have recently generated large interest, with examples of studies in both molecular qubits\cite{escalera2017determining,albino2019first,lunghi2019phonons,mirzoyan2020dynamic,lunghi2020limit,Kazmierczak2021} and single-ion magnets\cite{lunghi2017role,lunghi2017intra,goodwin2017molecular,moseley2018spin,ullah2019silico,lunghi2020multiple,reta2021ab,Ullah2021spectro,briganti2021}. Indeed, coordination compounds offer a versatile playground to test relaxation theories as their chemical structure and properties can be finely controlled and characterized experimentally\cite{zabala2021single,moreno2021measuring}. At the same time they allow for the use of advanced electronic structure methods to describe their lattice and magnetic properties\cite{atanasov2015first,ungur2017ab,neese2019chemistry}. 

However, until now, a fully quantitative prediction of relaxation times have been elusive and deviations of orders of magnitude with respect to experimental data have been observed. Such a situation casts shadows on our understanding of spin-phonon relaxation and on the reliability of \textit{ab initio} methods for blind prediction of spin relaxation time in absence of experimental validation.

In this contribution we advance the field of \textit{ab initio} spin dynamics by demonstrating that quantitative predictions of spin relaxation rate can be obtained for realistic solid-state compounds. We review the derivation of the dynamical equations for the spin reduced density matrix in solid-state materials\cite{lunghi2019phonons,lunghi2020limit,lunghi2020multiple} and individuate the key importance of including all the terms up to fourth-order perturbation theory and accounting for the dynamics of the entire density matrix, \textit{i.e.} including the explicit description of coherence terms' dynamics. We perform simulations for three systems: \textbf{(1)} an $S=1/2$ molecular qubit [VO(dmit)$_{2}$]$^{2-}$ (dmit=1,3-dithiole-2-thione-4,5-dithiolate)\cite{atzori2016quantum}, \textbf{(2)} an $S=3/2$ mononuclear complex [CoL$_{2}$]$^{2-}$ (H$_{2}$L = 1,2-bis(methanesulfonamido)benzene)\cite{Rechkemmer2016four,Yvonnephdthesis}, and \textbf{(3)} a $J=15/2$ mononuclear complex [DyCp$^{\textrm{ttt}}_2$]$^{+}$ (Cp$^{\textrm{ttt}}$=[C$_{5}$H$_{5} ^{\;\;\textrm{t}}$Bu$_{3}$-1,2,4])\cite{goodwin2017molecular}. The molecular structures of \textbf{(1)}-\textbf{(3)} are reported in Fig. \ref{geos}. These compounds represent the state-of-the-art in molecular magnetism for what concerns molecular qubits and single-molecule magnets with long spin lifetime. Moreover, their properties span a large range of magnetic splitting and relaxation times, providing an extremely robust proof-of-concept of the accuracy of \textit{ab initio} spin dynamics.\\

\section*{Results}

\textit{\textbf{Ab initio theory of spin-phonon relaxation.}} \\

The total system of spin plus phonon is described by the Hamiltonian
\begin{equation}
 \hat{H} =\hat{H}_{\mathrm{s}}+\hat{H}_{\mathrm{ph}}+\hat{H}_{\mathrm{s-ph}}\,
 \label{Htot}
\end{equation}
where the first two terms correspond to the spin and phonons Hamiltonians, respectively, and the third one represents the coupling between the two subsystems. Under a weak-coupling assumption, the interaction between a spin system and an ensemble of phonons, $q_{\alpha}$, can be modelled with a spin-phonon coupling Hamiltonian of the form,
\begin{align}
\hat{H}_{\mathrm{s-ph}} = & \sum_{\alpha} \left( \frac{\partial \hat{H}_{\mathrm{s}}}{\partial q_{\alpha}} \right)_{0} q_{\alpha}(t) \label{linsph} \\ 
+ & \sum_{\alpha \ge\beta} \left( \frac{\partial^{2} \hat{H}_{\mathrm{s}}}{\partial q_{\alpha}\partial q_{\beta}} \right)_{0} q_{\alpha}(t)q_{\beta}(t) \label{quadsph} \:,
\end{align}
where the summations are understood to run over both phonon band indexes and reciprocal lattice $\mathbf{q}$-points. The first- and second-order derivatives of $\hat{H}_{\mathrm{s}}$ are generally called spin-phonon coupling coefficients. The use of the spin Hamiltonian formalism in Eqs. \ref{Htot}, \ref{linsph}, and \ref{quadsph} does not constitute an approximation as long as an exact mapping between the low-lying electronic states of the system and the eigenstates of the spin Hamiltonian exists, which is generally the case\cite{atanasov2015first,ungur2017ab}.\\

The formalism of time-convolutionless master equations provides an exact description of the time-evolution of the spin reduced density matrix in the interaction picture\cite{breuer2001time,timm2011time}, $\hat{\rho}_s$, \textit{i.e.} once the explicit dynamics of the phonon degrees of freedom have been integrated out from the total Liouville equation
\begin{equation}
    \frac{d\hat{\rho}_{\mathrm{s}}}{dt}=\hat{\hat{\mathbf{R}}}(t)\hat{\rho}_{\mathrm{s}}(t)\:.
    \label{tcl}
\end{equation}
The relaxation super-operator $\hat{\hat{\mathbf{R}}}(t)$ can be systematically expanded in a perturbative way\cite{breuer2001time,timm2011time},  $\hat{\hat{\mathbf{R}}}(t)=(\hat{\hat{\mathbf{R}}}\mathbf{2}(t)+\hat{\hat{\mathbf{R}}}\mathbf{4}(t)+...)$, and in principles it can be implemented exactly if all the terms of Eq. \ref{Htot} are known. Assuming a separation of timescales between spin relaxation and the intrinsic relaxation rates of the phonons, mostly due to anharmonic phonon-phonon scattering, it is possible to assume that the latter are always at thermal equilibrium and therefore apply the Born-Markov approximation. Namely, we assume that phonons have a much larger specific heat than the spin system (no phonon bottleneck\cite{tesi2016giant}) and that they are able to rapidly equilibrate while exchanging energy with the spin. Under these assumptions and considering only the second-order contribution $\hat{\hat{\mathbf{R}}}\mathbf{2}(t)$, Eq. \ref{tcl} becomes identical to the well-known Redfield equations\cite{lunghi2019phonons}
\begin{equation}
\frac{d\rho^{\mathrm{s}}_{ab}(t)}{dt}=\sum_{cd}e^{i(\omega_{ac}+\omega_{db})t}R2^{n-\mathrm{ph}}_{ab,cd}\rho^{\mathrm{s}}_{cd}(t)\:.
\label{redfield}
\end{equation}

Considering only the linear term of Eq. \ref{linsph}, $\hat{\hat{\mathbf{R}}}\mathbf{2}^{n-\mathrm{ph}}$ becomes
 \begin{widetext}
 \begin{align}
 R2^{1-\mathrm{ph}}_{ab,cd} =-\frac{\pi}{2\hbar^{2}} & \sum_{\alpha}\Big\{\sum_{j} \delta_{bd}V^{\alpha}_{aj}V^{\alpha}_{jc} G^{1-\mathrm{ph}}(\omega_{jc},\omega_{\alpha}) 
 -V^{\alpha}_{ac}V^{\alpha}_{db}G^{1-\mathrm{ph}}(\omega_{bd},\omega_{\alpha}) \nonumber \\
 & -V^{\alpha}_{ac}V^{\alpha}_{db}G^{1-\mathrm{ph}}(\omega_{ac},\omega_{\alpha})  +\sum_{j}\delta_{ca}V^{\alpha}_{dj}V^{\alpha}_{jb}G^{1-\mathrm{ph}}(\omega_{jd},\omega_{\alpha})\Big\} \label{Red21} \:,
\end{align}
\end{widetext}
where the terms $V^{\alpha}_{ab}$ are a short-hand notation for $\langle a | (\partial \hat{H}_{\mathrm{s}}/\partial q_{\alpha})  | b \rangle$, and $| a \rangle$ and $| b \rangle$ are eigenstates of $\hat{H}_{\mathrm{s}}$. Finally, $\omega_{ab}=(E_{a}-E_{b})/\hbar$, where $E_{a}$ and $E_{b}$ are eigenvalues of $\hat{H}_{\mathrm{s}}$. The function $G^\mathrm{1-ph}$ that appears in Eq. \ref{Red21} is the Fourier transform of the single-phonon correlation function and reads
\begin{equation}
G^\mathrm{1-\mathrm{ph}}(\omega,\omega_{\alpha})=\delta(\omega-\omega_{\alpha})\bar{n}_{\alpha}+\delta(\omega+\omega_{\alpha})(\bar{n}_{\alpha}+1)\:,
\label{G1ph}
\end{equation}
where $\bar{n}_{\alpha}=[\mathrm{exp}(\hbar\omega_{\alpha}/\mathrm{k_\mathrm{B}}T)-1]^{-1}$ is the Bose-Einstein distribution of thermal population, $\hbar\omega_{\alpha}$ is the $\alpha$-phonon energy and $\mathrm{k_B}$ is the Boltzmann constant. The function $G^{1-\mathrm{ph}}$ accounts for the spectral density and population of phonons. Eqs. \ref{redfield} and \ref{Red21}, describe one-phonon resonant spin transitions, such as those responsible for Direct and Orbach relaxation. Considering now the quadratic term of Eq. \ref{quadsph}, $\hat{\hat{\mathbf{R}}}\mathbf{2}^{n-\mathrm{ph}}$ becomes\cite{lunghi2020limit}
\begin{widetext}
\begin{align}
R2^{2-\mathrm{ph}}_{ab,cd} =-\frac{\pi}{4\hbar^{2}} & \sum_{\alpha\ge\beta} \Big\{\sum_{j} \delta_{bd}V^{\alpha\beta}_{aj}V^{\alpha\beta}_{jc} G^{2-\mathrm{ph}}(\omega_{jc},\omega_{\alpha},\omega_{\beta})-V^{\alpha\beta}_{ac}V^{\alpha\beta}_{db}G^{2-\mathrm{ph}}(\omega_{bd},\omega_{\alpha},\omega_{\beta}) \nonumber \\
&- V^{\alpha\beta}_{ac}V^{\alpha\beta}_{db}G^{2-\mathrm{ph}}(\omega_{ac},\omega_{\alpha},\omega_{\beta})+\sum_{j}\delta_{ca}V^{\alpha\beta}_{dj}V^{\alpha\beta}_{jb}G^{2-\mathrm{ph}}(\omega_{jd},\omega_{\alpha},\omega_{\beta})\Big\}\:, \label{Red22}
\end{align}
\end{widetext}
where $V^{\alpha\beta}_{ab}$ now stands for $\langle a | (\partial^{2} \hat{H}_{\mathrm{s}}/\partial q_{\alpha}\partial q_{\beta})  | b \rangle$. It should be noted that the term $\alpha=\beta$ is only included for double absorption/emission terms (\textit{vide infra}). The function $G^{2-\mathrm{ph}}$ accounts for three possible processes involving two phonons: absorption of two phonons, emission of two phonons and simultaneous emission of one phonon and absorption of another one
\begin{widetext}
\begin{align}
    G^{2-\mathrm{ph}}(\omega_{ba},\omega_{\alpha},\omega_{\beta}) = & \delta(\omega_{ba}-\omega_{\beta}+\omega_{\alpha})\bar{n}_{\beta}(\bar{n}_{\alpha}+1) + 
     \delta(\omega_{ba}+\omega_{\beta}-\omega_{\alpha})(\bar{n}_{\beta}+1)\bar{n}_{\alpha} + \\
     & \delta(\omega_{ba}-\omega_{\beta}-\omega_{\alpha})\bar{n}_{\beta}\bar{n}_{\alpha} + 
      \delta(\omega_{ba}+\omega_{\beta}+\omega_{\alpha})(\bar{n}_{\beta}+1)(\bar{n}_{\alpha}+1)\:.
    \label{G2sph}
\end{align}
\end{widetext}
As evident from Eq. \ref{G2sph}, in two-phonon transitions, only the sum or difference of the two phonons' energies must be resonant with the spin transition, while the energy of the single vibrations could be anything. Spin relaxation mechanisms due to two-phonon processes generally falls under the definition of Raman relaxation mechanism. \\

At this point of the derivation of the Redfield equations, the secular approximation is generally performed. This approximation involves setting to zero the matrix elements of $R2^{n-\mathrm{ph}}_{ab,cd}$ for which $(\omega_{ac}+\omega_{db})\ne 0$. This is justified by the fact that the terms $(\omega_{ac}+\omega_{db})$ appear as arguments of an oscillatory term in Eq. \ref{redfield}. These oscillating terms are averaged out for times commensurable to the spin relaxation time, $\tau$, provided that the period of the natural spin oscillations ($\omega_{ab}$) is much shorter than $\tau$. Bearing in mind that spin-phonon relaxation in magnetic molecules often occurs on timescales of ns to hundreds of seconds, depending on the value of $T$ and the specific system, the secular approximation is generally fulfilled for systems with zero-field splitting or in the presence of small external or dipolar magnetic fields. Under these conditions, only the terms $R2^{n-\mathrm{ph}}_{aa,bb}$ and $R2^{n-\mathrm{ph}}_{ab,ab}$, namely population transfer and coherence relaxation terms, contribute to the dynamics of $\hat{\rho}_{s}$. However, when dealing with Kramers systems in zero external magnetic field, additional terms of $R2^{n-\mathrm{ph}}_{ab,cd}$ also survive the secular approximation in virtue of the spectrum degeneracy that makes additional terms $(\omega_{ac}+\omega_{db})$ vanish. For instance, the population-coherence transfer process $R2^{n-\mathrm{ph}}_{aa,bc}$, with $E_{b}=E_{c}$, and coherence transfer terms $R2^{n-\mathrm{ph}}_{ad,bc}$, with $E_{b}=E_{c}$ and $E_{a}=E_{d}$, would not be washed out by the secular approximation and could in principles contribute to the spin-phonon dynamics\cite{eastham2016bath,gosweiner2020spin}. \\

Extending the expansion of $\hat{\hat{\mathbf{R}}}(t)$ up to the fourth-order and considering only the linear term of spin-phonon coupling, we obtain another source of two-phonon spin transitions. A set of equations describing the diagonal population terms of $\hat{\rho}_{s}$ due to this contribution reads\cite{timm2011time}
\begin{equation}
\frac{d\rho^{\mathrm{s}}_{aa}(t)}{dt}=\sum_{bb}R4^{2-\mathrm{ph}}_{aa,bb}\rho^{\mathrm{s}}_{bb}(t)\:,
\end{equation}
where
\begin{widetext}
\begin{align}
& R4^{2-\mathrm{ph}}_{aa,bb}=  \frac{\pi}{2\hbar^{2}} \sum_{\alpha \ge\beta} \left [ A_{\alpha \beta} W^{--}_{ab}(\alpha \beta) + A_{\alpha \beta} W^{++}_{ab}(\alpha \beta) + B_{\alpha \beta} W^{+-}_{ab}(\alpha \beta) + B_{\alpha \beta} W^{-+}_{ab}(\alpha \beta) \right] \:, \: \text{and}\\ 
  &   W^{--}_{ab}(\alpha \beta) =  \Big|\sum_{c} \frac{\langle a |\hat{V}_{\alpha}|c\rangle\langle c|\hat{V}_{\beta} | b \rangle}{E_{c}-E_{b}-\hbar\omega_{\beta}}+
    \frac{\langle a |\hat{V}_{\beta}|c\rangle\langle c|\hat{V}_{\alpha} | b \rangle}{E_{c}-E_{b}-\hbar\omega_{\alpha}} \Big|^{2} \bar{n}_{\alpha} \bar{n}_{\beta} \delta(\omega_{ab}-\omega_{\alpha}-\omega_{\beta})  \\
  &  W^{++}_{ab}(\alpha \beta) =   \Big|\sum_{c}
    \frac{\langle a |\hat{V}_{\alpha}|c\rangle\langle c|\hat{V}_{\beta} | b \rangle}{E_{c}-E_{b}+\hbar\omega_{\beta}}+
    \frac{\langle a |\hat{V}_{\beta}|c\rangle\langle c|\hat{V}_{\alpha} | b \rangle}{E_{c}-E_{b}+\hbar\omega_{\alpha}} \Big|^{2} (\bar{n}_{\alpha}+1) (\bar{n}_{\beta}+1) \delta(\omega_{ab}+\omega_{\alpha}+\omega_{\beta})  \\
  &  W^{+-}_{ab}(\alpha \beta) =   \Big|\sum_{c}
    \frac{\langle a |\hat{V}_{\alpha}|c\rangle\langle c|\hat{V}_{\beta} | b \rangle}{E_{c}-E_{b}+\hbar\omega_{\beta}}+
    \frac{\langle a |\hat{V}_{\beta}|c\rangle\langle c|\hat{V}_{\alpha} | b \rangle}{E_{c}-E_{b}-\hbar\omega_{\alpha}} \Big|^{2} \bar{n}_{\alpha} (\bar{n}_{\beta}+1) \delta(\omega_{ab}-\omega_{\alpha}+\omega_{\beta})  \\
  &   W^{-+}_{ab}(\alpha \beta) =   \Big|\sum_{c}
    \frac{\langle a |\hat{V}_{\alpha}|c\rangle\langle c|\hat{V}_{\beta} | b \rangle}{E_{c}-E_{b}-\hbar\omega_{\beta}}+
    \frac{\langle a |\hat{V}_{\beta}|c\rangle\langle c|\hat{V}_{\alpha} | b \rangle}{E_{c}-E_{b}+\hbar\omega_{\alpha}} \Big|^{2} (\bar{n}_{\alpha}+1) \bar{n}_{\beta} \delta(\omega_{ab}+\omega_{\alpha}-\omega_{\beta})  \:,
    \label{Red41}
\end{align}
\end{widetext}
where $A_{\alpha \beta}=(1-3/4\delta_{\alpha\beta})$ and $B_{\alpha \beta}=(1-1/2\delta_{\alpha\beta})$. One can recognize in Eq. \ref{Red41} the same contributions of $G^{2-ph}$, except for different pre-factors, which are now accounting for the presence of excited spin states, \textit{i.e.} the virtual states often invoked in perturbation theory. The processes describe by Eq. \ref{Red41} also contributes to the so-called Raman relaxation. It is important to remark that Eq. \ref{Red41} should also include additional terms that cancel the divergence of the denominators\cite{koller2010density,timm2011time}. We neglect those terms in light of the fact that the numerical sparsity of the phonon spectrum never leads to such divergences. Importantly, Eq. \ref{Red41} differs from a previously presented one, where not all the terms had been included\cite{lunghi2020multiple,briganti2021}. Finally we note that Eq. \ref{Red41} does not account for the dynamics of coherence terms and is therefore only strictly valid for non-degenerate spin spectra. \\

As long as the Born-Markov and secular approximations hold, the equations just detailed provide an exact description of spin relaxation up to two-phonon processes. The only remaining challenge to their implementation lies in the definition of the many coefficients that enter these equations. This challenge is tackled with electronic structure methods. Methods such as Complete Active Space Self Consistent Field (CASSCF) and Density Functional Theory (DFT) have now reached a high degree of sophistication in the prediction of spin Hamiltonian parameters of magnetic molecules\cite{atanasov2015first,ungur2017ab} as well to predict phonons' frequencies and normal modes of vibrations\cite{garlatti2020unveiling}. \\

\textit{\textbf{Spin-phonon relaxation in $S=1/2$ systems.}}\\

\begin{figure}
 \begin{center}
    \includegraphics[scale=1]{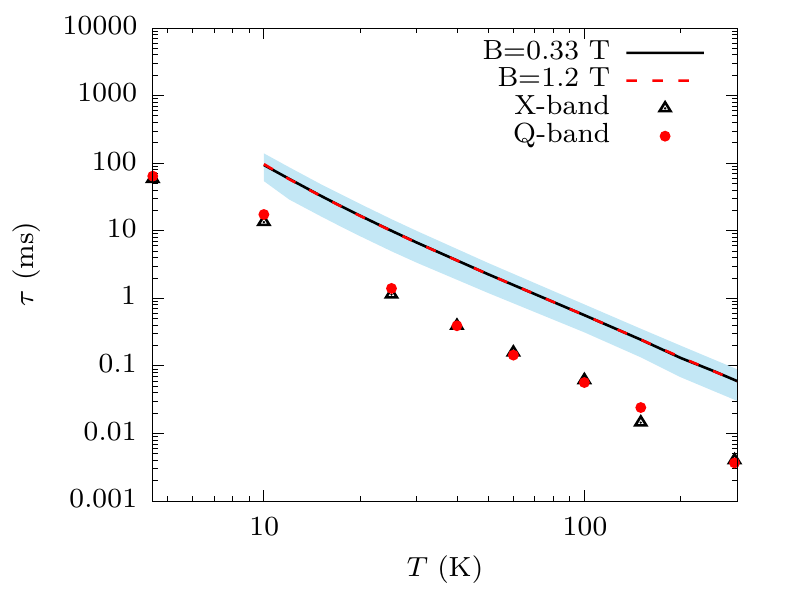} \\
    \includegraphics[scale=1]{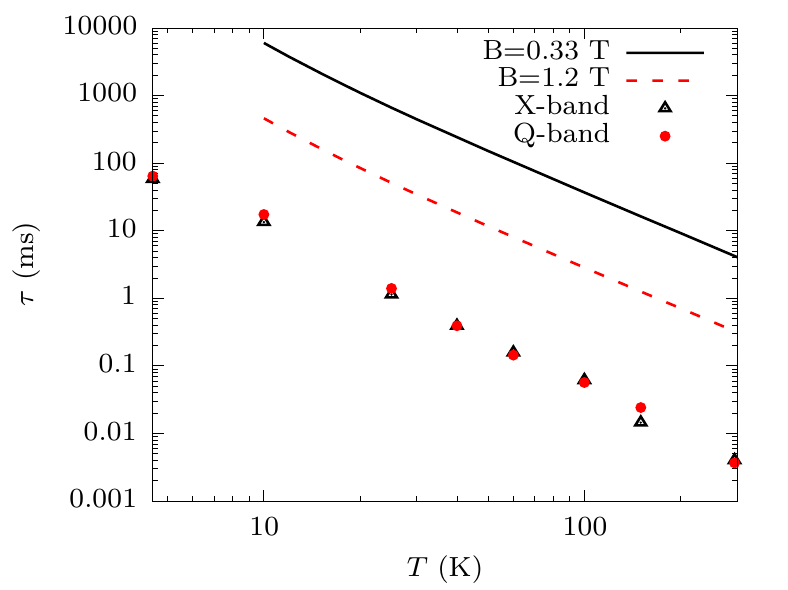} 
\end{center}
 \caption{\textbf{Spin-phonon relaxation time for \textbf{(1)}.} Simulated values of $\tau$ for \textbf{(1)} including the contribution of \textbf{A}-tensor (top panel) and \textbf{g}-tensor (bottom panel) modulations to the Raman mechanism of relaxation. Results in the top panel are the average between calculations carried out with the DKH and ZORA scalar relativistic corrections and the sky-blue shade represents the two limiting values. Experimental values correspond to the T$_{1}$ times obtained from inversion recovery EPR experiments at X- and Q-bands.}
 \label{vorate}
\end{figure}

Compound \textbf{(1)} is a V$^{4+}$ coordination complex with an $S=1/2$ ground state and a nuclear spin $I=7/2$\cite{atzori2016quantum}. The spin Hamiltonian can be modelled as
\begin{equation}
\hat{H}_{\mathrm{s}}=\mu_{\mathrm{B}}\vec{\mathbf{B}}\cdot \mathbf{g} \cdot \vec{\mathbf{S}} + \gamma_{\mathrm{N}}\vec{\mathbf{B}}\cdot \mathbf{\vec{I}} + \vec{\mathbf{S}}\cdot \mathbf{A} \cdot \vec{\mathbf{I}} \:,
\label{vosph}
\end{equation}
where $\mu_{\mathrm{B}}$ is the electron's Bohr magneton, $\mathbf{g}$ is the effective electronic Landè tensor, $\gamma_{\mathrm{N}}$ is the nuclear gyromagnetic factor of $^{51}$V of and $\mathbf{A}$ is the hyperfine coupling tensor. \textbf{(1)} crystallizes in a monoclinic unit-cell containing four molecular units and eight tetraphenylphosphonium  counter-ions. The entire unit-cell was optimized with periodic-DFT (pDFT) and used for phonons calculations as detailed in the Computational Methods. The tensors  $\mathbf{g}$ and $\mathbf{A}$ were computed with DFT after a structural optimization. Predicted values of $(g_{xx},g_{yy},g_{zz})=(1.976,1.986,1.988)$ are in perfect agreement with experimental ones\cite{atzori2016quantum}. Computed values of $\mathbf{A}$ are found sensitive to the choice of scalar relativistic corrections used in the DFT simulation, with ZORA underestimating the values up to 23\% and DKH overestimating them up to 38\%. All the terms of the spin-phonon coupling Hamiltonian of \textbf{(1)}, including second order terms, are computed by taking the numerical derivative of all the parameters in Eq. \ref{vosph} that depend on molecular coordinates and affect spin dynamics, namely $\mathbf{g}$ and $\mathbf{A}$. Notably, here we perform the calculation of second-order spin-phonon coupling coefficients fully \textit{ab initio} and without using any machine-learning interpolator as done in previous work\cite{lunghi2020limit}. We neglect the contributions of Eq. \ref{Red21} to relaxation, which are known to lead to Direct relaxation\cite{lunghi2019phonons}. The latter only overcome two-phonon relaxation for $T<10$ K\cite{atzori2016quantum,lunghi2020limit}. In that regime experimental values of $\tau$ are affected by spin diffusion, which would make the comparison with simulations not straightforward. Eqs. \ref{redfield} and \ref{Red22} are solved for an initial condition of the spin density matrix where a $\pi$ rotation of the electronic spin is applied to the canonical equilibrium distribution. This initial state is chosen to mimic the experimental condition of an inversion recovery electron paramagnetic resonance (EPR) experiment used to measure T$_{1}$. The time-dependence of the magnetization of the electronic spin is computed to reach equilibrium following a stretched-exponential decay $M_{z}(t)=(M_{z}(0)-M_{z}^{eq})$ exp$[-(t/\tau)^{\beta}]+M_{z}^{eq}$ with $\beta\sim 0.7-0.9$ for $A$-driven relaxation and $\beta=1$ for $g$-driven relaxation. Fig. \ref{vorate} reports the comparison between the values of T$_{1}$ measured with X- and Q-Band EPR on magnetically diluted samples, together with the predicted values of $\tau$ as a function of temperature by assuming the modulation of $\mathbf{A}$ (top panel) and $\mathbf{g}$ (bottom panel) as the sole relaxation mechanisms.
\begin{figure*}
 \begin{center}
    \includegraphics[scale=0.85]{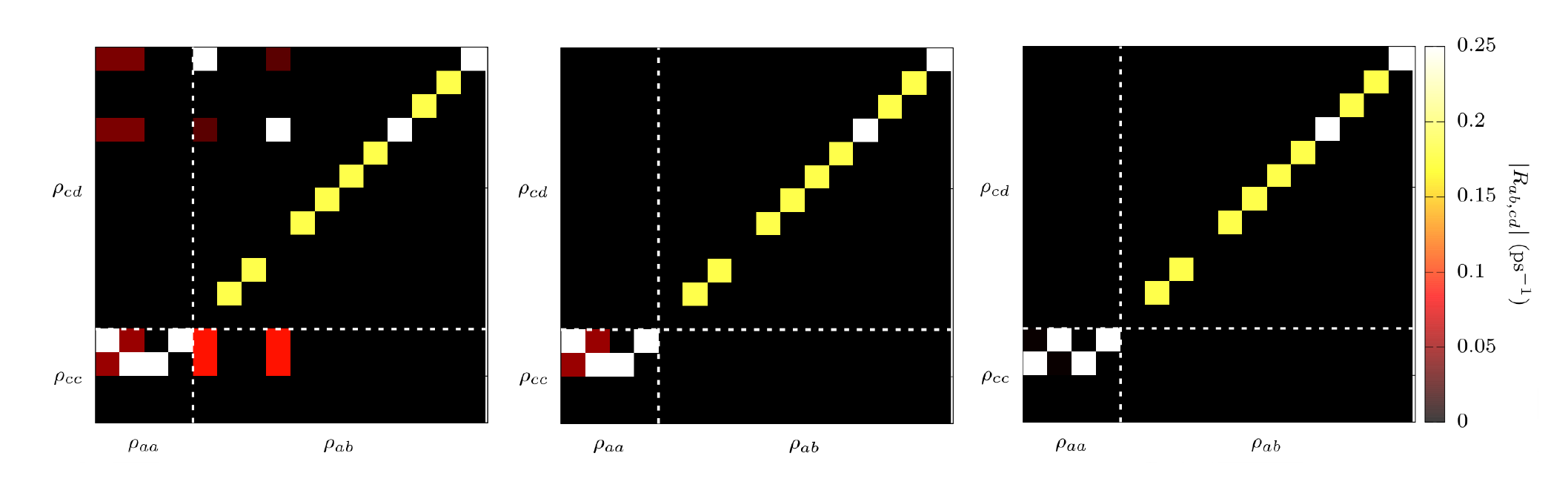}
\end{center}
 \caption{\textbf{Redfield Transition Rates for \textbf{(2)}.} Transition rates among general elements of the density matrix predicted by the Redfield theory employing the non-diagonal secular approximation and diagonal secular approximation are reported in the left and central panel, respectively. The right panel reports the same rates for the non-diagonal secular approximation after rotating the molecular geometry and applying a small external field. The transitions rates have been computed at $T=40$ K. The first four elements correspond to the population terms, and all the remaining corresponds to coherence terms. }
 \label{maps}
\end{figure*}
Results obtained considering the modulation of the hyperfine interaction show a good agreement with inversion recovery experiments, both in terms of absolute values and external field dependence. Depending on the fine details of the DFT calculations agreement up to a factor five and not exceeding one order of magnitude is observed. Conversely, the modulation of the Zeeman interaction is shown to overestimate spin lifetime by at least one order of magnitude. More importantly, the relaxation rate due to the modulation of \textbf{g} is found to depend on the magnitude of the external field as $B^{2}$, which does not fit the generally observed experimental trend for Raman relaxation in this class of complexes. The computational results for \textbf{(1)} are in qualitative agreement with a previous study on the crystal of VO(acac)$_{2}$\cite{lunghi2020limit}. However, differently from the VO(acac)$_{2}$ case, T$_{1}$ data obtained on diluted crystals for \textbf{(1)} are available\cite{atzori2016quantum}, making it possible to demonstrate that once cross-relaxation mediated by dipolar interactions is excluded, the deviation between simulations and experiments is below one order of magnitude and the $\tau$ vs $T$ profile is nicely reproduced. Indeed, the residual deviation is consistent with the lack of phonons outside the $\Gamma$-point and the inaccuracies of electronic structure methods. In order to determine the most important phonons for relaxation, we perform simulations by including only vibrations in the low-energy window and starting from the first optical mode at $\Gamma$-point, here computed at $\sim 12$ cm$^{-1}$. We find that optical phonons up to $\sim 50$ cm$^{-1}$ are the main responsible for spin relaxation even at room temperature. This can be interpreted as an effect of thermal population, which is always larger for lowest-energy phonons. As discussed previously, low-energy vibrations of \textbf{1} are an admixture of molecular rotations and delocalized intra-molecular distortions\cite{albino2019first}. \\

\textit{\textbf{Spin-phonon relaxation in $S>1/2$ systems.}} \\

Next we address the spin dynamics of \textbf{(2)} as an example of relaxation in a mononuclear transition-metal complex with large zero-field splitting. This molecule contains a high-spin Co$^{2+}$ ion that can be described with an effective $S=3/2$ Hamiltonian

\begin{equation}
\hat{H}_{\mathrm{s}}=D\hat{S}_{z} + E(\hat{S}_{x}-\hat{S}_{y}) \:,
\label{CoSH}
\end{equation}
where the large axial anisotropy term $D\sim -115$ cm$^{-1}$ removes the degeneracy of the two Kramers doublets (KDs) $M_{s}=\pm 3/2$ and $M_{s}=\pm 1/2$\cite{Rechkemmer2016four,Yvonnephdthesis}. This compound crystallizes in an orthorhombic lattice with four molecules in the unit cell and eight NHEt$_{3}^{+}$ counter-ion  molecules\cite{Rechkemmer2016four}. All the spin-phonon coupling coefficients and the phonons for this compound were computed in a previous work employing CASSCF and DFT-based machine-learning force fields (MLFF), respectively\cite{lunghi2020multiple}. In this previous work, the time evolution of $\hat{\rho}^{s}$ was studied with the diagonal terms of Eq. \ref{Red21} and Eq. \ref{Red41}, therefore neglecting the role of coherence terms. In order to investigate the relevance of these terms we compute all the terms of $R2^{1-\mathrm{ph}}_{ab,cd}$. Fig. \ref{maps}a shows the value of $R2^{1-\mathrm{ph}}_{ab,cd}$ when only the terms $R2^{1-\mathrm{ph}}_{aa,bb}$ and $R2^{1-\mathrm{ph}}_{ab,ab}$ are retained. We will refer to this approximation as diagonal secular approximation from now on. Instead, Fig. \ref{maps}b shows the value of $R2^{1-\mathrm{ph}}_{ab,cd}$ when all terms that fulfil the condition $(\omega_{ac}+\omega_{db})=0$ are retained. We will refer to this level of approximation as non-diagonal secular approximation from now on. As it is evident from the comparison of Figs. \ref{maps}a-b, the non-diagonal secular approximation includes non-negligible terms that couples coherence and population elements of $\rho^{\mathrm{s}}_{ab}$. Finally, Fig. \ref{maps}c shows that the diagonal secular approximation becomes fulfilled when Kramers degeneracy is removed by applying a small external field of 0.01 Tesla along the molecule's easy axis. We note that the eigenvalues of the matrices reported in Figs. \ref{maps}a,c are virtually identical, and at the same time different from those of the matrix reported in Fig. \ref{maps}b. Moreover, we find that the spectrum of $R$ in the diagonal secular approximation is not rotationally invariant even in zero-field, a clear indication of inconsistency. 

\begin{figure}
 \begin{center}
    \includegraphics[scale=1]{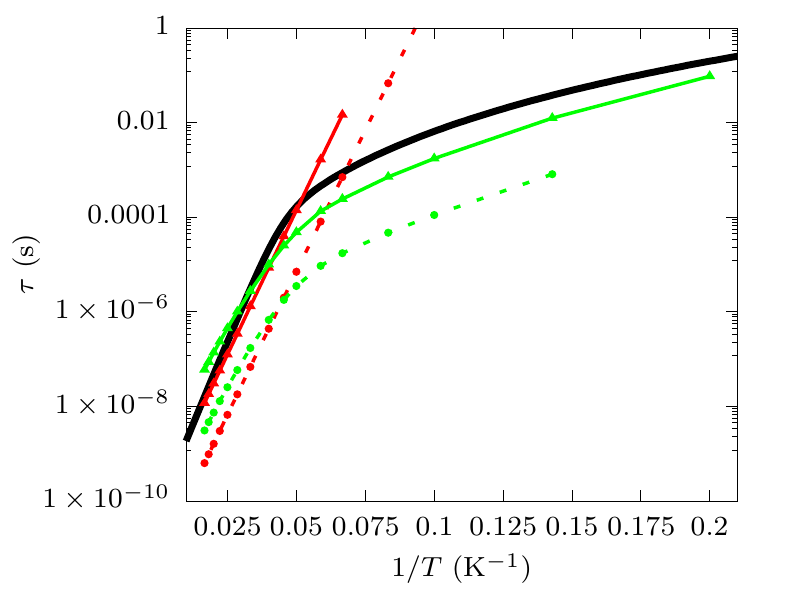} 
\end{center}
 \caption{\textbf{Spin-phonon relaxation time for \textbf{(2)}.} Experimental relaxation times measured at 1000 Oe with AC magnetometry are reported as black continuous lines\cite{Yvonnephdthesis}. Simulated Orbach rates computed with the non-diagonal (continuous line and triangles) and diagonal secular approximation (dashed line and circles) are reported in red. Simulated Raman rates computed with the diagonal secular approximation with external field $B_{z}=0.01$ Tesla (continuous line and triangles) and $B_{z}=0.0$ Tesla (dashed line and circles) are reported in green. Raman rates in non-zero external field are computed by rotating the molecular geometry in the frame of the g-tensor's eigenvectors.}
 \label{Corates}
\end{figure}

In order to determine the importance of including coherence-transfer terms into the description of spin-phonon relaxation, we compute the value of spin lifetime $\tau$ for \textbf{(2)} by taking the inverse of the first non-zero eigenvalue of the matrix $R2^{1-\mathrm{ph}}_{ab,cd}$, as commonly done for Markov processes. The latter is equivalent to studying the decay of $M_{z}(t)$ for a molecule with the magnetization easy-axis parallel to $z$. Fig. \ref{Corates} shows the results for the two different levels of secular approximation, together with experimental results obtained in the presence of a small external field. The application of a small external field in experiments helps removing the effect of dipolar relaxation\cite{Yvonnephdthesis}, not included in the simulations and therefore makes the comparison between theory and experiments more reliable. The relaxation time predicted with the diagonal secular approximation is the same reported before\cite{lunghi2020multiple} and despite the qualitative agreement with experiments, it shows between one and two orders of magnitude of underestimation of spin lifetime. We note that the prediction of Raman contributions with Eq. \ref{Red41} only slightly improves with respect to those presented previously\cite{lunghi2020multiple}. Once the non-diagonal secular approximation is introduced for one-phonon processes with the use of Eq. \ref{Red21}, the agreement between experimental and simulated $\tau$ becomes virtually exact in the high-$T$ relaxation regime, where the Orbach mechanism drives relaxation. For temperature below 20 K, the two-phonon Raman relaxation described by Eq. \ref{Red41} takes over the Orbach one. Although a non-diagonal secular approximation for this process is not available, we attempted to remove the effect of coherence transfer by applying a small external field along the molecular easy-axis of magnetization. As suggested by the study of $R2^{1-\mathrm{ph}}_{ab,cd}$ matrix elements, the results for the diagonal and non-diagonal secular approximations becomes identical under these conditions. Applying this strategy to $R4_{aa,bb}^{2-\mathrm{ph}}$, results improve significantly with only a negligible residual deviation left between experiments and simulations. Finally, we note that simulations for \textbf{(2)} have been found extremely robust with respect to the choice of basis set, scalar relativistic effects, and dynamical correlation. Moreover, predictions obtained with $\Gamma-$point phonons computed from MLFF or DFT agree very well among them, thus validating the use of MLFF\cite{lunghi2019unified} to integrate the phonons' Brillouin zone and converge the values of $\tau$ below 10 K, where border-zone and acoustic phonons start contributing. As discussed elsewhere\cite{lunghi2020multiple} and in agreement with what is observed here for \textbf{(1)}, low energy phonons up to $\sim 50$ cm$^{-1}$ are the main drive for Raman spin relaxation in \textbf{(2)}. Also in this case, it was found that these low energy vibrations correspond to molecular rotations combined to small and delocalized intra-molecular distortions\cite{lunghi2020multiple}. \\

\textit{\textbf{Spin-phonon relaxation in J=15/2 lanthanide systems.}} \\

\begin{figure}
 \begin{center}
    \includegraphics[scale=1]{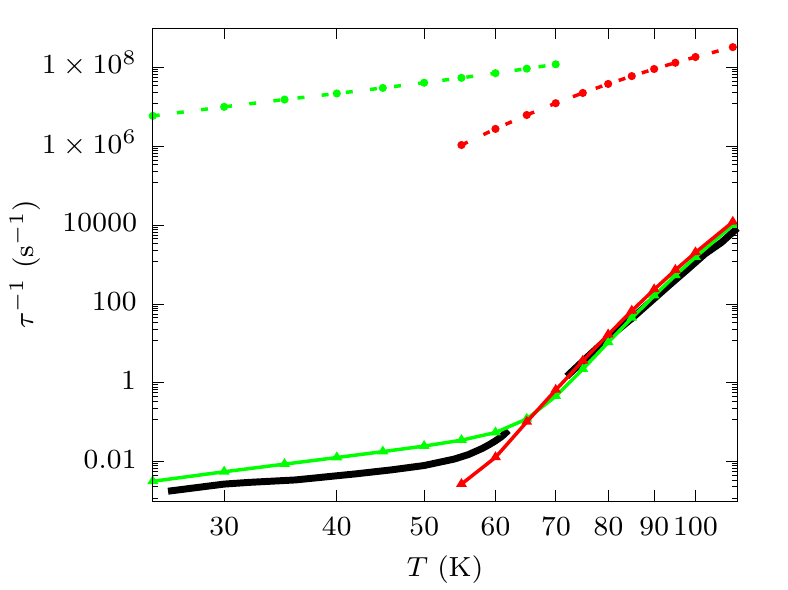}
\end{center}
 \caption{\textbf{Spin-phonon relaxation time for \textbf{(3)}.} Experimental relaxation times measured in zero field with AC magnetometry and magnetization decay are reported as a black continuous line\cite{goodwin2017molecular}. Simulated Orbach rates computed with the non-diagonal (continuous line and triangles) and diagonal secular approximation (dashed line and circles) are reported in red. Simulated Raman rates computed with the diagonal secular approximation with external field $B_{z}=0.01$ Tesla (continuous line and triangles) and $B_{z}=0.0$ Tesla (dashed line and circles) are reported in green. Raman rates in the presence of non-zero external field are computed by rotating the molecular geometry in the frame of the g-tensor's eigenvectors of the ground-state KD.}
 \label{Dyrates}
\end{figure}

In order to further test the ability of our models, we extended our study to \textbf{(3)}. This molecule exhibits a Dy$^{3+}$ ion in almost perfectly axial crystal field. This coordination stabilizes a $J=15/2$ ground state and imposes a strong zero-field splitting that separates the ground- and highest excited KDs ($Mj=\pm 15/2$ and $Mj=\pm 1/2$, respectively) by $\sim 1500$ cm$^{-1}$\cite{goodwin2017molecular}. The spin states of \textbf{(3)} can be described with a generalized spin Hamiltonian\cite{ungur2017ab,jung2019derivation}
\begin{equation}
\hat{H}_{s}=\sum_{l=2,4,6}\sum_{m=-l}^{l}=B^{l}_{m}\hat{O}_{m}^{l}(\vec{\mathbf{J}})\:,
\label{CFH}
\end{equation}
where $\hat{O}_{m}^{l}(\vec{\mathbf{J}})$ are tesseral tensor operators\cite{tennant2000rotation}. \textbf{(3)} crystallizes in a triclinic unit-cell with two molecular units as well as two  [B(C$_{6}$F$_{5}$)$_{4}$]$^{-}$\cite{goodwin2017molecular} molecules as counter-ions. Phonons and spin-phonon coupling coefficients were computed for \textbf{(3)} following the strategy validated on \textbf{(2)} and that involves the computation of $\Gamma-$point phonons with pDFT and linear spin-phonon coupling coefficients with CASSCF. Results of spin relaxation for the one-phonon Orbach process in \textbf{(3)} are reported in Fig. \ref{Dyrates} together with experimental values. Once again, the agreement between the simulated and experimental Orbach relaxation rates is excellent, but only after the  non-diagonal secular approximation is introduced. The prediction of Orbach rates obtained using the diagonal secular approximation is off by many orders of magnitude, depending on the value of $T$. Such a large disagreement is mostly due to the fact that in absence of the non-diagonal secular approximation, relaxation is predicted to be promoted by the first excited KD, computed at $\sim$450 cm$^{-1}$, instead of being driven by absorption of phonons in resonance with the KD with energy $\sim 1300$ cm$^{-1}$. We tested the effect of removing Kramers degeneracy by aligning the Cartesian frame along the molecular easy-axis, as expressed by the eigenvectors of ground-state KD's g-tensor, and by applying an external field of 0.01 Tesla along the same direction. In agreement with what observed for \textbf{(2)}, under these conditions, Orbach relaxation rates predicted with the diagonal and the non-diagonal secular approximations are coincident among them and with experimental relaxation rates. Exporting our computational strategy to the case of Raman relaxation, we observe a massive improvement of results, with a negligible factor of residual deviation between experiments and simulations. Differently from \textbf{(2)}, we note a slight dependence of results with respect to basis set, the number of CASSCF's solutions considered and the inclusion of dynamical correlation, hinting to the possibility to further improve results upon a systematic exploration of these technicalities. 
Turning to the analysis of phonon contributions to Raman spin relaxation in \textbf{(3)}, relaxation was found to receive contributions from several optical phonons with energy up to 80 cm$^{-1}$. A visual inspection of these modes shows that similarly to \textbf{(2)} they are dominated by local molecular rotations overlapped to small intra-molecular distortions. The latter mainly involve the $^{\mathrm{t}}$Bu groups and the position of Dy$^{3+}$ ion within the two Cp$^{-}$ rings. 
We determined the contribution of different excited KDs to Eq. \ref{Red41} by including only one of them at the time in the simulation of $\tau$ at 40 K. This analysis shows that Raman relaxation is almost entirely determined by the contribution of the first excited KD. \\

\section*{Discussion} 

Although the theory of relaxation for $d$/$f$-block ions has been thoroughly investigated for decades, the many parameters populating its equations have prevented its use as a predictive tool. However, once relaxation theory is combined with \textit{ab initio} methods, it becomes possible to estimate spin relaxation time without any input from experiments. Such a method stands out as a key tool to benchmark our understanding of spin-phonon interaction and relaxation in an unbiased way and to provide new strategies towards optimization of relaxation times. Nonetheless, a computational method of such complexity possesses many assumptions and approximations that require careful validation.

In this work we have shown that a careful comparison with experimental data, a consistent use of all fourth-order processes, and the inclusion of coherence terms in the Redfield equations are key for obtaining quantitative predictions. Most importantly we have shown how seemingly small details, such as neglecting the coherence terms of the density matrix, can lead to many orders of magnitude of deviation between predictions and experiments. As long as a parametric approach to open-quantum-system dynamics is used, it is impossible to detect such pitfalls, as errors are removed by rescaling factors. These results thus provide a unique proof of concept that open quantum system theory can be applied in a parameter-free fashion for complex systems such as solid-state.

We envisage that further improvement of the agreement between experiments and simulations could be achieved by carefully benchmarking both electronic structure methods and the remaining theoretical approximations to the open-quantum system dynamical equations. The former includes a careful study of how vdW corrections to DFT affect phonons' simulations and how much electronic structure correlation and the choice of basis-set impact the prediction of spin Hamiltonian and spin-phonon coupling coefficients. From a theoretical point of view, the derivation of an equation able to predict the evolution of the entire spin density matrix up to the fourth-order stands out as the next fundamental challenge. Assessing the validity of the Markov approximation in situations of phonon-bottleneck\cite{tesi2016giant} and the inclusion of spin-spin interactions into the formalism are two other interesting directions for the field.  

The unprecedented level of accuracy of simulations for such different compounds validates the theoretical framework and the \textit{ab initio} methods at the same time, making it possible to confirm recent theoretical findings in terms of spin relaxation mechanism in vdW crystals of molecular Kramers system\cite{lunghi2019phonons,chiesa2020understanding,lunghi2020limit,gu2020origins,lunghi2020multiple,briganti2021}. Direct and Orbach relaxation due to resonant phonons are correctly accounted for by second-order contributions to density matrix dynamics and linear spin-coupling. Two-phonon spin relaxation due to second-order contributions to density matrix dynamics and quadratic spin-phonon coupling is the main source of Raman relaxation for $S=1/2$, while fourth-order perturbation theory and linear coupling is necessary to explain Raman relaxation in $S>1/2$ with large uni-axial anisotropy. Interestingly, our results show that only THz vibrations contribute to Raman relaxation, regardless of the energy of excited KDs. Simulations thus support the use of the expression 
\begin{equation}
    \tau^{-1} = \sum_{i} \frac{A^{1-\mathrm{ph}}_{i} }{(e^{\beta \omega_{i}}-1)} + \sum_{i} A^{2-\mathrm{ph}}_{i} \frac{e^{\beta \omega_{i}} }{ (e^{\beta \omega_{i}}-1)^{2}} 
    \label{fiteq}
\end{equation}
to fit one- and two-phonon contributions to experimental spin lifetime of vdW crystals of magnetic molecules\cite{lunghi2020limit}. The first term of Eq. \ref{fiteq} can either lead to Direct or Orbach relaxation depending on the energy of resonant phonons. This approach has been used already on a few occasions, providing additional insights with respect to a fitting done with simple power-laws\cite{santanni2020probing,de2021exploring,Kazmierczak2021,pfleger2021terminal}. \\

In conclusion, we have studied spin-phonon relaxation in three molecular compounds representing the most relevant classes of slow-relaxing Kramers systems and shown that open-quantum system theory can be implemented in a fully-\textit{ab initio} fashion and used to accurately predict spin relaxation rates. Now that \textit{ab initio} spin dynamics is rapidly maturing into a quantitative predictive tool it will be possible to deploy it to study other systems such as poly-nuclear ion clusters\cite{feltham2014review}, organic radicals\cite{eaton2002relaxation}, nuclear spins\cite{corzilius2014dynamic}, paramagnetic defects in semiconductors\cite{jarmola2012temperature} and magnetic impurities in solid-state hosts\cite{wolfowicz2021quantum} or adsorbed on surfaces\cite{donati2016magnetic}. \\

\section*{Computational Methods}

\textit{\textbf{Electronic structure simulations.}} The unit-cell X-ray structure of \textbf{(1)}\cite{atzori2016quantum} and \textbf{(3)}\cite{goodwin2017molecular} were used as starting points for a periodic DFT optimization with the software CP2K\cite{kuhne2020cp2k}. Density functional theory (DFT) with the PBE functional\cite{perdew1996generalized}, including Grimme's D3 van der Waals corrections~\cite{grimme2010consistent}, was used together with a double-zeta polarised (DZVP) MOLOPT basis set. A plane-wave cutoff of 2500 Ry, and 1500 Ry was used for \textbf{(1)} and \textbf{(3)}, respectively. All the unit-cell lattice force constants were computed with a two-point numerical differentiation with step of 0.01 \AA. Geometry optimization, phonons and generation of a MLFF for \textbf{(2)} were presented before. The ORCA software~\cite{neese2020orca} has been employed for the calculation of all spin Hamiltonian terms. Simulations with ORCA were carried out on molecular geometries optimized with pDFT. The tensors \textbf{g} and \textbf{A} for \textbf{(1)} were computed with DFT employing the PBE0 functional\cite{perdew1996rationale}. The spin Hamiltonian of \textbf{(3)} was computed with CASSCF employing a (9,7) active space and by using all the solutions with multiplicity six, 128 solutions with multiplicity four, and 130 solutions with multiplicity two. Spin-orbit contributions were included through quasi-degenerate perturbation theory. ZORA and DKH scalar relativistic corrections were used for \textbf{(1)} and \textbf{(3)}, respectively. The calculation of spin-phonon coupling coefficients for \textbf{(1)} was also repeated with DKH scalar relativistic corrections. The RIJCOSX approximation with GridX6 integration grid was used for both \textbf{(1)} and \textbf{(3)}. The basis sets DKH/ZORA-def2-TZVPP was used for V, S, C and H. SARC-DKH-TZVPP was used for Dy instead. Spin-phonon coupling coefficients for \textbf{(3)} were computed using a slightly cheaper setup: state-average CASSCF simulations only included the roots with multiplicity six, and RIJCOSX was used with the GridX4 option. The basis sets DKH-SARC-def2-QZVP, DKH-def2-TZVPP and DKH-def2-SVP were used for Dy, C, and H, respectively. Spin Hamiltonian and spin-phonon coupling coefficients for \textbf{(2)} were computed previously with CASSCF\cite{lunghi2020multiple}. \\

\textit{\textbf{Spin-phonon relaxation simulations.}} Phonons' frequency and normal modes of vibrations were computed by diagonalization of the dynamical matrix 
\begin{equation}
    D_{ij}(\mathbf{q})= \sum_l \Phi^{0,l}_{ij} e^{i\mathbf{q}\cdot\mathbf{R}_l}\:,
\end{equation}
where $\mathbf{q}$ is a Brillouin zone vector and $\Phi^{0,l}_{ij}$ is the mass-weighed lattice force constant between the degree of freedom $i$ in the reference unit-cell $l=0$ and the degree of freedom $j$ in the unit-cell $l$. First-order spin-phonon coupling coefficients are computed with the expression
\begin{equation}   
\Big(\frac{\partial \hat{H}_\mathrm{s}}{\partial q_{\alpha\mathbf{q}}}\Big)=\sum_{i}^{3N}\sqrt{\frac{\hbar}{N_{q}\omega_{\alpha\mathbf{q}}m_{i}}} L^{\mathbf{q}}_{\alpha i} \Big(\frac{\partial \hat{H}_\mathrm{s}}{\partial X_{i}}\Big)\:,
\label{sph}
\end{equation}
where $L^{\mathbf{q}}_{\alpha i}$ and $\omega_{\alpha\mathbf{q}}^{2}$ are the eigenvectors and eigenvalues of $D_{ij}(\mathbf{q})$, respectively, $N_q$ is the total number of q-points used to integrate the Brillouin zone and the sum over $i$ is extended to the $3N$ molecular degrees of freedom. A similar expression is used for second-order spin-phonon coupling\cite{lunghi2020limit}. The derivatives of the spin Hamiltonian terms appearing in Eq. \ref{sph} are computed by numerical differentiation. In the case of $(\partial B^{l}_{m}/\partial X_i)$, the parameters $B^{l}_{m}$ were computed for six distortions between $\pm$ 0.05 \AA$ $ for each molecular degree of freedom. Their value was then interpolated with a third-order polynomial function, where the linear coefficient corresponds to $(\partial B^{l}_{m}/\partial X_i)$. For the second-order differentiation of g and A, we used a four-point finite-difference expression with step 0.01 \AA. Dirac's delta functions appearing in the spin dynamics equations are smeared out with a Gaussian function with $\sigma=10$ cm$^{-1}$. A value of $i\sigma$ is also added to the denominators of Eq. \ref{Red41} to prevent possible divergence of calculated rates. Results are found to be well converged with respect to $\sigma$ according to a strategy previously tested\cite{lunghi2020limit,lunghi2020multiple}. Low-temperature results for Raman relaxation in \textbf{(2)} were also converged by sampling the vibrational Brillouin zone thanks to supercell phonons calculations\cite{lunghi2020multiple}.

\vspace{0.2cm}
\noindent
\textbf{Acknowledgements}\\
This project has received funding from the European Research Council (ERC) under the European Union’s Horizon 2020 research and innovation programme (grant agreement No. [948493]). Computational resources were provided by the Trinity College Research IT and the Irish Centre for High-End Computing (ICHEC).

\vspace{0.2cm}
\noindent
\textbf{Conflict of interests}\\
The authors declare that they have no competing interests.


%

\end{document}